\documentclass[sigconf]{acmart}

\AtBeginDocument{%
  }

\setcopyright{acmlicensed}
\copyrightyear{2026}
\acmYear{2026}
\setcopyright{cc}
\setcctype{by}
\acmConference[DAC '26]{63rd ACM/IEEE Design Automation Conference}{July 26--29, 2026}{Long Beach, CA, USA}
\acmBooktitle{63rd ACM/IEEE Design Automation Conference (DAC '26), July 26--29, 2026, Long Beach, CA, USA}
\acmDOI{10.1145/3770743.3804417}
\acmISBN{979-8-4007-2254-7/2026/07}

\usepackage{algorithmic}
\usepackage{graphicx}
\usepackage{textcomp}
\usepackage[dvipsnames]{xcolor}
\usepackage{multirow}
\usepackage{makecell}
\usepackage{enumitem}
\usepackage{booktabs}
\usepackage{url}
\usepackage{hyperref}
\usepackage{amsthm}
\usepackage[ruled]{algorithm2e}
\usepackage{setspace}
\usepackage{comment}
\usepackage{subcaption}
\usepackage[utf8]{inputenc}

\newcommand{\probP}{\text{I\kern-0.15em P}}
\newlist{researchquestions}{enumerate}{1}
\setlist[researchquestions]{label*=\textbf{RQ\arabic*}}

\DeclareMathOperator*{\argmin}{arg\,min}

\begin{document}

\title{Scalable and Adaptive Parallel Training of Graph Transformer on {Large Graphs}}

\author{Jun-Liang Lin}
\affiliation{%
  \institution{The Pennsylvania State University}
  \city{University Park}
  \state{PA}
  \country{USA}
}
\email{jpl6521@psu.edu}

\author{Kamesh Madduri}
\affiliation{%
  \institution{The Pennsylvania State University}
  \city{University Park}
  \state{PA}
  \country{USA}
}
\email{madduri@psu.edu}

\author{Mahmut Taylan Kandemir}
\affiliation{%
  \institution{The Pennsylvania State University}
  \city{University Park}
  \state{PA}
  \country{USA}
}
\email{mtk2@psu.edu}

\begin{abstract}
Graph foundation models have demonstrated remarkable adaptability across diverse downstream tasks through large-scale pretraining on graphs. However, existing implementations of the backbone model, graph transformers, are typically limited to single-GPU systems, leading to long training times or out-of-memory issues on {large graphs}. Moreover, parallelizing graph transformer training over the full graph is challenging, as efficiency depends heavily on both the graph structure and system characteristics, such as bandwidth and memory capacity.

In this work, we introduce a distributed training framework for graph transformers, which automatically selects and optimizes parallelization strategies based on the graph structure and hardware configuration. With our implementation of distributed sparse operations, we accelerate sparse graph attention by up to 3.8x and reduce memory consumption by 78\% compared to state-of-the-art frameworks.  On {large graph} benchmarks, our proposed framework achieves up to 6x speedup with system scaling up to 8 GPUs. These results demonstrate that the proposed framework improves the scalability of graph transformers, bringing them closer to serving as practical graph foundation models.
\end{abstract}

\maketitle
\vspace{-0.1in}
\section{Introduction \label{sec:intro}}
With the rapid development of transformer models, research has shown that scaling up both the size of the model and the pretraining data leads to stronger performance~\cite{hoffmann2022training}. Larger pretrained models exhibit remarkable adaptability across a wide range of downstream tasks, demonstrating strong transferability and generalization~\cite{achiam2023gpt, dubey2024llama, comanici2025gemini}. These pretrained models are commonly referred to as “foundation models” because they serve as a universal base (foundation) on which diverse applications can be built, allowing performance improvements by leveraging the rich knowledge encoded within them.

Extending this paradigm to graph-structured data, graph foundation models aim to learn universal representations through large-scale self-supervised pretraining, enabling broad applicability to downstream tasks via finetuning or prompt tuning~\cite{gui2024g, sun2022gppt, liu2024one}. Among candidate architectures, “graph transformers” have emerged as a promising backbone due to their ability to capture both structural and semantic patterns in graphs~\cite{dwivedi2020generalization, shi2021masked, rampavsek2022recipe, shirzad2023exphormer, Zhuo2025}. Recent studies have explored their effectiveness in multiple domains and various pretraining strategies have been proposed to further enhance their generalization and transferability~\cite{gui2024g, zi2024prog}.

Scaling graph transformers to larger graphs is crucial for advancing graph foundation models~\cite{liu2025graph}. However, recent graph transformer designs mainly target single-GPU training~\cite{dwivedi2020generalization, shi2021masked, rampavsek2022recipe, shirzad2023exphormer}, and therefore fail to scale algorithms and models to {large graphs. In contrast, modern graph neural network (GNN) training frameworks, such as DistDGL~\cite{zheng2020distdgl} and BGL~\cite{285052}, do support distributed multi-GPU training, but they are mostly designed for mini-batch training, where subgraphs are sampled and processed in each iteration.} This paradigm is not applicable to general graph transformers, which rely on full-graph information to capture long-range dependencies. {Although some distributed training frameworks support full-graph GNN workloads, they are largely optimized for simple GCN models~\cite{wan2022bns,wan2023scalable,10.14778/3725688.3725700}. None of these frameworks support graph transformers, which involve substantially more complex computation and communication patterns.}

{Several recent approaches, such as SGFormer~\cite{wu2023simplifying} and TorchGT~\cite{10.1109/SC41406.2024.00083}, attempt to scale graph transformers through graph clustering or partitioning, but they inevitably sacrifice parts of the graph structure, losing the ability to fully exploit global information. In other words, no existing framework provides an efficient solution for distributed graph transformer training with full graphs.} These limitations motivate a critical question: \textit{How can we enable distributed training of graph transformer models on full graphs while ensuring scalability on multi-GPU systems?}

Following the transformer design used in large language models (LLMs), general graph transformers incorporate query, key, and value projections along with a self-attention mechanism. However, a key distinction is that instead of computing full attention on all pairs of nodes, graph transformers restrict attention to neighboring nodes, resulting in a sparse attention mask~\cite{dwivedi2020generalization, rampavsek2022recipe, shirzad2023exphormer, shi2021masked}. This mask corresponds to the adjacency matrix of the graph and can be represented in sparse formats such as Coordinate (COO) or Compressed Sparse Row (CSR)~\cite{saad1990sparskit}. Consequently, attention computation is performed through a series of sparse operations, including sparse matrix–matrix multiplication (SpMM) and sampled dense–dense matrix multiplication (SDDMM)~\cite{8638042, rahman2021fusedmm}. To achieve scalable training of graph transformers, it is crucial to distribute these sparse operations efficiently across multiple devices.

Unlike LLMs, where model parameters dominate memory consumption, training graph transformers is often constrained by the size of the graph data itself. As a result, traditional parallelization strategies used in LLM training, such as data parallelism and model parallelism~\cite{narayanan2021efficient}, are not effective for scaling graph transformer training on {large graphs}. Therefore, we focus on “graph parallelism”, referring to exploiting parallelism based on the graph structure and its features. In this direction, there exist multiple strategies, each with different computation costs, communication overhead, and memory footprints~\cite{10.14778/3717755.3717776}. Determining the optimal strategy is challenging due to the diversity of graph structures and the variability in system configurations.

In this paper, we address these challenges by introducing a ``distributed training framework'' to enable training of the graph transformer on full {large graphs}. The main {\bf contributions} of this work are as follows: 

$\bullet$ We propose two parallelization strategies, Graph Parallelism with All-Gather (GP-AG) and Graph Parallelism with All-to-All (GP-A2A), to enhance training efficiency through the use of sparse operators and reduce memory consumption significantly.

$\bullet$ We introduce an algorithm, Automatic Graph Parallelism (AGP), which automatically applies the optimal parallelization strategy to training of sparse graph transformers on {large graphs}, based on the target graph structure and system configurations. 

$\bullet$  We present a detailed experimental evaluation of our algorithm. This evaluation demonstrates strong scalability (achieving up to a 6x speedup on 8 GPUs) for full-graph training on {large graphs}, along with substantial improvements in training efficiency over prior approaches (up to a 3.8x speedup and a 78\% reduction in memory usage). 

\section{Graph Transformers with Sparse Graph Attention~\label{sec:sparse}}

\subsection{Definition of Sparse Graph Transformer}
A typical graph transformer design follows the definition in the early work UniMP~\cite{shi2021masked}, which can be represented as follows: 

\begin{equation} \label{eq:unimp} 
    \mathbf{x}^{\prime}_i = \mathbf{W}_o \mathbf{x}_i +
        \sum_{j \in \mathcal{N}(i)} \alpha_{i,j} \mathbf{W}_V \mathbf{x}_{j},
\end{equation}

\begin{equation} 
    \alpha_{i,j} = \textrm{softmax} \left(
        \frac{(\mathbf{W}_Q\mathbf{x}_i)^{\top} (\mathbf{W}_K\mathbf{x}_j)}
        {\sqrt{d}} \right),
\end{equation}
\vspace{0.1in}

\noindent where $\mathbf{x}_i$ and $\mathbf{x}^{\prime}_i$ are the input and output features of node $i$, $\mathbf{W}$s are the weight matrices, and $\mathcal{N}(i)$ is the set of neighbors of node $i$. Note that we can rewrite and extend it into the matrix form. Given a graph $G$ with $N$ nodes and feature dimension $d$, along with a dense feature matrix $\textbf{X} \in \mathbb{R}^{N\times d}$ and a sparse adjacency matrix $\textbf{A} \in \{0,1\}^{N \times N}$, the general form of the ``Sparse Graph Attention'' (SGA) can be calculated as follows:

\begin{equation} \label{eq:qkv}  
   \textbf{Q}=\textbf{X}\textbf{W}_Q, 
   \textbf{K}=\textbf{X}\textbf{W}_K, 
   \textbf{V}=\textbf{X}\textbf{W}_V,
\end{equation}

\begin{equation} \label{eq:sga}  
    \textbf{Z} = 
    (\textbf{Q}\textbf{K}^\top) \odot \textbf{A}, \textbf{U} = \text{Softmax}(\frac{\textbf{Z}}{\sqrt{d}}),
\end{equation}

\begin{equation} \label{eq:uv}  
    \textbf{Y} = \textbf{U}\textbf{V}, \textbf{X'} = \textbf{X}\textbf{W}_o + \textbf{Y},
\end{equation}

\noindent where $\textbf{W}_Q$, $\textbf{W}_K$, and $\textbf{W}_V$ are the weight matrices of size $d \times d$, $\textbf{Z}$ and $\textbf{U}$ are sparse matrices, and $\textbf{Y}$ is the summation term in Equation~\ref{eq:unimp}. The attention mechanism can also be extended into multi-head attention (MHA), which splits the features into multiple subspaces, computes attention independently in each, and then combines the results to capture diverse relational patterns in parallel.

\subsection{Sparse Operations During {Forward and Backward Propagation}}
For large graphs, the adjacency matrix $\textbf{A}$ is usually represented in a sparse format such as COO or CSR. Consequently, computing the sparse graph attention requires several sparse operations. Specifically, the computation of $(\textbf{Q}\textbf{K}^\top) \odot \textbf{A}$ in Equation~\ref{eq:sga} can be efficiently implemented with SDDMM~\cite{8638042, rahman2021fusedmm}. Note that the output of this operation is also a sparse matrix. Thus, multiplication with matrix $\textbf{V}$ in Equation~\ref{eq:uv} can be performed using SpMM. As a result, the entire forward propagation of SGA involves 1 SDDMM, 1 SpMM, and the 3 dense matrix multiplications (MM) in Equation~\ref{eq:qkv}. {For backward propagation, each SpMM in the forward pass requires one SpMM and one SDDMM in the backward pass, while each SDDMM in the forward pass requires two SpMM operations in the backward pass. Therefore, the backward pass requires a total of three SpMM operations and one SDDMM operation.}


\begin{figure}[t]
  \centering
  \includegraphics[width=0.98\linewidth]{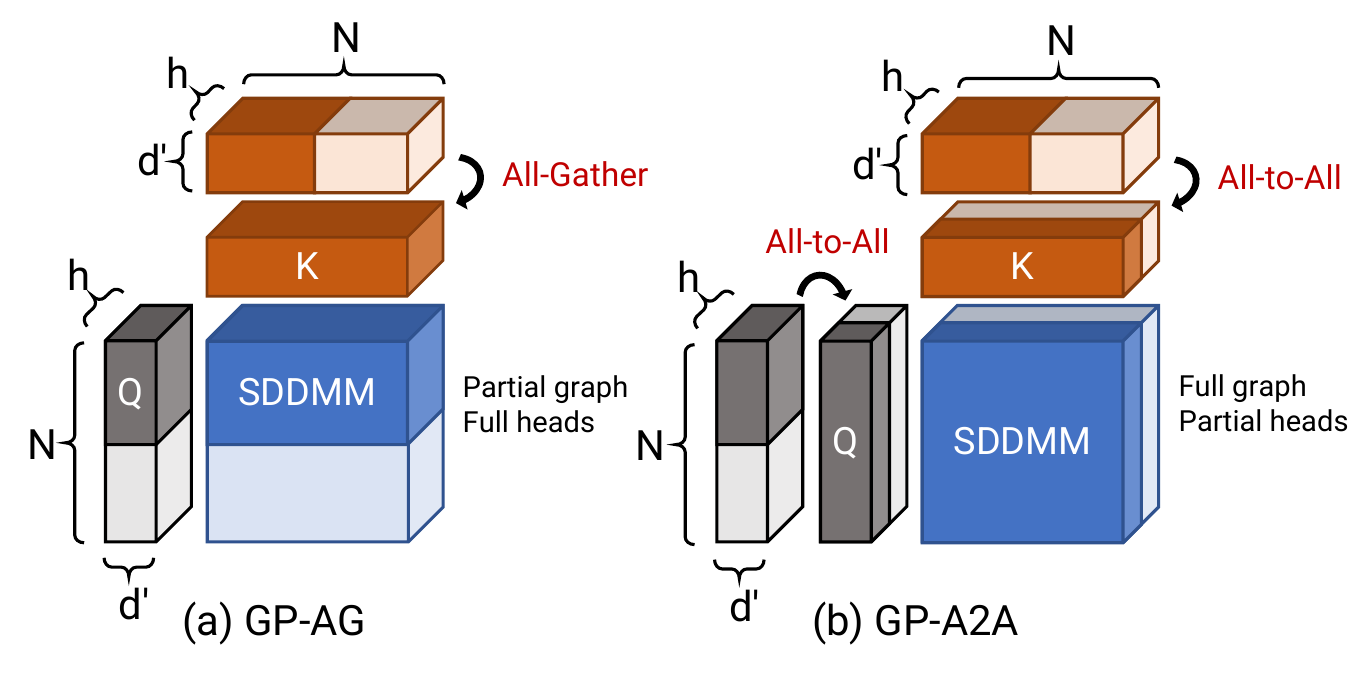}
  \caption{The two different parallelization strategies for calculating the attention matrix. $N$ denotes the number of nodes, $h$ the number of heads, and $d'$ the head dimension.  \label{fig:gp}}
\end{figure}

\section{Distributed Training of Sparse Graph Transformers} 
In this section, we discuss two parallelization strategies in terms of communication cost and peak memory consumption. The first strategy, {\bf GP-AG} (Graph Parallelism with All-Gather), distributes nodes across workers and invokes all-gather operations~\cite{chan2007collective} during the forward propagation of sparse graph attention. The second strategy, {\bf GP-A2A} (Graph Parallelism with All-to-All), also distributes nodes across workers but invokes all-to-all communication~\cite{chan2007collective} during forward propagation. Figure~\ref{fig:gp} illustrates the process of collecting matrices Q and K and the corresponding calculation of the attention matrix with these two parallelization strategies. 

\subsection{GP-AG: Graph Parallelism with All-Gather}
Consider training graph transformers using $p$ GPUs, where nodes are partitioned across workers. In this setting, the dense matrix multiplications required for the \textbf{QKV} transformations incur no additional communication overhead (Equation~\ref{eq:qkv}). When calculating the attention matrix in Equation~\ref{eq:sga}, each device must gather the entire matrix \textbf{K}, causing a data transfer of $\mathcal{O} ( N \cdot d \cdot \frac{p-1}{p} )$. For Equation~\ref{eq:uv}, the entire matrix \textbf{V} also needs to be gathered, causing another $\mathcal{O} ( N \cdot d \cdot \frac{p-1}{p} )$ data transfer. \textit{For clarity, we simplify the term $\frac{p-1}{p}$ in the following discussion.} Algorithm~\ref{alg:gp_ag_fw} outlines the detailed steps of this process.  The all-gather communication performed in the forward pass necessitates a matching Reduce-Scatter operation in the backward pass. Overall, GP-AG incurs two all-gather and two reduce-scatter calls for the attention computation, while each worker computes the attention over its assigned subgraph for all attention heads.

\begin{algorithm}[t]
\caption{Forward propagation of GP-AG.}
\label{alg:gp_ag_fw}
\small
\LinesNumbered
\KwIn{$X$, $A$, $W_Q$, $W_K$, $W_V$ }
\KwOut{$Y$}
$Q$, $K$, $V$ $\gets$ $XW_Q$, $XW_K$, $XW_V$;
\textcolor{Blue}{$K_{all}$ $\gets$ all-gather ($K$); }  \tcc{[(N/p),h,d']->[N,h,d']}
\textcolor{Brown}{$Z$ $\gets$ SDDMM($A$, $Q$, $K_{all}^\top$);}\\
$U$ $\gets$ softmax($Z$/$\sqrt{d}$);\\
\textcolor{Blue}{$V_{all}$ $\gets$ all-gather ($V$);}\\
\textcolor{Brown}{$Y$ $\gets$ SpMM($U$, $V_{all}$);} \tcc{[(N/p),h,d']}
\Return{$Y$}
\end{algorithm}

\begin{algorithm}[t]
\caption{Forward propagation of GP-A2A.}
\label{alg:gp_a2a_fw}
\small
\LinesNumbered
\KwIn{$X$, $A$, $W_Q$, $W_K$, $W_V$ }
\KwOut{$Y$}
$Q$, $K$, $V$ $\gets$ $XW_Q$, $XW_K$, $XW_V$;
\textcolor{Blue}{$Q'$ $\gets$ all-to-all ($Q$);} \tcc{[(N/p),h,d']->[N,(h/p),d']}
\textcolor{Blue}{$K'$ $\gets$ all-to-all ($K$);}\\
\textcolor{Brown}{$Z'$ $\gets$ SDDMM($A$, $Q'$, $K'^\top$);}\\
$U'$ $\gets$ Softmax($Z'$/$\sqrt{d}$);\\
\textcolor{Blue}{$V'$ $\gets$ all-to-all ($V$);}\\
\textcolor{Brown}{$Y'$ $\gets$ SpMM($U'$, $V'$);}\\
\textcolor{Blue}{$Y$ $\gets$ all-to-all ($Y'$);} \tcc{[N,(h/p),d']->[(N/p),h,d']}
\Return{$Y$}
\end{algorithm}
               
\subsection{GP-A2A: Graph Parallelism with All-to-All}
Since the calculations for each attention head are independent, they do not introduce additional communication overhead. With this in mind, we employ an all-to-all strategy to efficiently distribute the computational workload. In the GP-A2A training, each worker computes the \textbf{Q}, \textbf{K}, and \textbf{V} matrices for a subset of nodes. We then use all-to-all communication to exchange data along the split dimension, transforming it from node-based partitioning (dimension $N$) to head-based partitioning (dimension $h$). This allows each worker to collect all nodes but only a portion of the features. Attention computation proceeds in parallel across workers, with each worker responsible for different attention heads. Finally, another all-to-all communication is used to revert the partitioning back to dimension $N$, restoring the initial partitioning scheme. Algorithm~\ref{alg:gp_a2a_fw} outlines the detailed steps of this process.  Each all-to-all communication in the forward pass results in a corresponding all-to-all operation in the backward pass. Consequently, GP-A2A incurs a total of eight all-to-all calls for the attention computation, with each worker computing attention over the entire graph but only for a portion of the attention heads. Table~\ref{table:communication_memory} provides a summary of the communication costs and memory consumption for the two strategies.

\section{Automatic Graph Parallelism for Sparse Graph Transformers}
In this section, we propose an {\em analytical model} to estimate training throughput based on graph properties and system configuration. We then explain how to determine the coefficients in the model and estimate the performance as the system scales. 

\begin{table}[t]
\centering
\small
\caption{Comparison of communication costs and memory consumption for different parallelization strategies per attention block per GPU. AG: All-Gather, RS: Reduce-Scatter, A2A: All-to-All.}
\begin{tabular}{c|ccccc}
\toprule
       & Comm. & Cost & Act. Memory & Graph storage  \\ 
\midrule
GP-AG   & 2 AG + 2 RS & 4$Nd$ & $4Nd+\frac{Eh}{p}$  & $\frac{N}{p}+\frac{E}{p}$  \\
GP-A2A  & 8 A2A & 8$\frac{Nd}{p}$  & $\frac{4Nd}{p}+\frac{Eh}{p}$   & $N+E$   \\
\bottomrule
\end{tabular}
\label{table:communication_memory}
\end{table}

\begin{table}[t]
\centering
\small
\caption{The properties of the graph datasets and the microbenchmark of compute kernels on an A100 GPU. }
\begin{tabular}{l|cc|ccc}

\toprule
\textbf{Dataset} & \textbf{Nodes} & \textbf{Edges}  & \textbf{MM} & \textbf{SpMM} & \textbf{SDDMM}  \\ 
\midrule
ogbn-arxiv & 169K & 1.1M & 0.33 ms & 0.97 ms & 0.45 ms \\
ogbn-proteins & 132K   & 79M & 0.27 ms   & 51.63 ms      & 25.11 ms                \\
ogbn-products & 2.4M & 123M  & 4.84 ms  & 91.43 ms     & 54.43 ms                 \\ 
reddit    & 233K   & 114M & 0.46 ms    & 76.55 ms       & 39.14 ms                 \\

\bottomrule
\end{tabular}
\label{tab:microbenchmark}
\end{table}

\subsection{Workload Analysis}
We begin by examining how the size of the graph impacts the training performance of the graph transformers by profiling the key kernels involved in training. Typically, three types of kernel are utilized. Dense matrix-multiplication (denoted as MM) is used to calculate the \textbf{Q}, \textbf{K}, and \textbf{V} matrices, with a problem size of $N \times d \times d$. On the other hand, SDDMM is used when calculating the attention matrix, with a problem size of $N \times d \times N$. Finally, SpMM is used when multiplying the attention matrix by the \textbf{V} matrix, with a problem size of $N \times N \times d$.

Table~\ref{tab:microbenchmark} presents the graph properties alongside the runtime performance of the SpMM, SDDMM, and MM kernels. From these experiments, we derive two key insights:

$\bullet$ The execution time of sparse operations is determined mainly by the number of edges, whereas the runtime of dense operations depends on the number of nodes.

$\bullet$ Sparse operations, particularly SpMM and SDDMM, dominate the computational workload throughout the training process.

Thus, it becomes essential to focus on analyzing and optimizing sparse operations to enhance the efficiency of graph transformer training. 

\subsection{Theoretical Runtime Analysis}
To estimate the throughput of graph transformers as the system scales, we begin by modeling the theoretical runtime. In distributed training, the total iteration time can be expressed as follows:  
\begin{equation} 
t_{iter} = t_{compute} + t_{comm}. 
\end{equation}
According to our previous analysis, the runtime of the compute kernels in a distributed graph transformer is related to the number of edges, and the runtime of communication kernels is related to the number of nodes. Therefore, we can rewrite the above equation as a function of the number of GPUs $p$ as follows:
\begin{equation}   
t_{iter}(p) = \alpha(p) \cdot E + \beta_c(p) \cdot N, 
\end{equation} 
\noindent where $\alpha(\cdot)$ is the coefficient that approximates the computation time based on the number of edges, and $\beta_c(\cdot)$ is the coefficient related to the bandwidth and the type of collective operation $c$, which is highly dependent on the system topology and the underlying implementations.

For {large graphs} or large problem sizes, $\alpha(\cdot)$ can be estimated as:
\begin{equation}   
\alpha(sp) \approx \frac{\alpha(p)}{s}, 
\end{equation}
where $s$ is the scaling factor. On the other hand, $\beta_c(\cdot)$ can be viewed as a property of the system and can be estimated using profiling tools such as NCCL Tests~\cite{nccl-tests}.

\begin{figure}
\centering
    \begin{subfigure}[b]{0.235\textwidth}
    \centering
    \includegraphics[width=\textwidth]{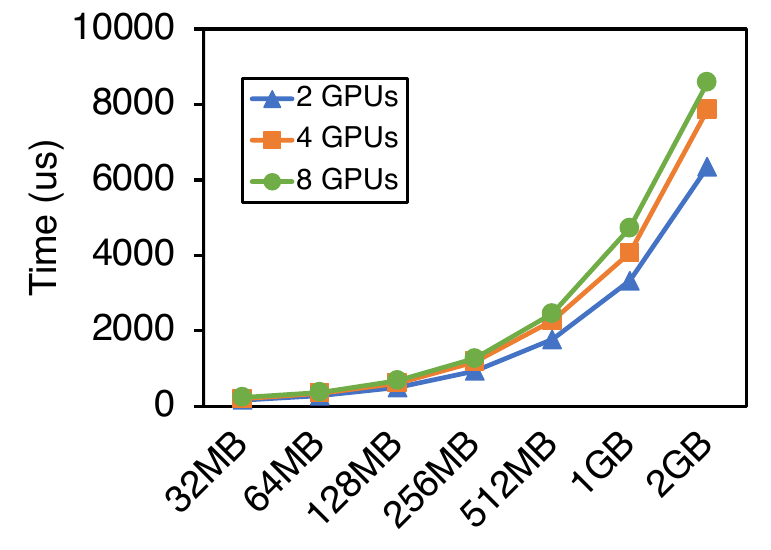}
    \caption{All-gather\label{fig:ag_test}}
    \end{subfigure}
    \begin{subfigure}[b]{0.235\textwidth}
    \centering
    \includegraphics[width=\textwidth]{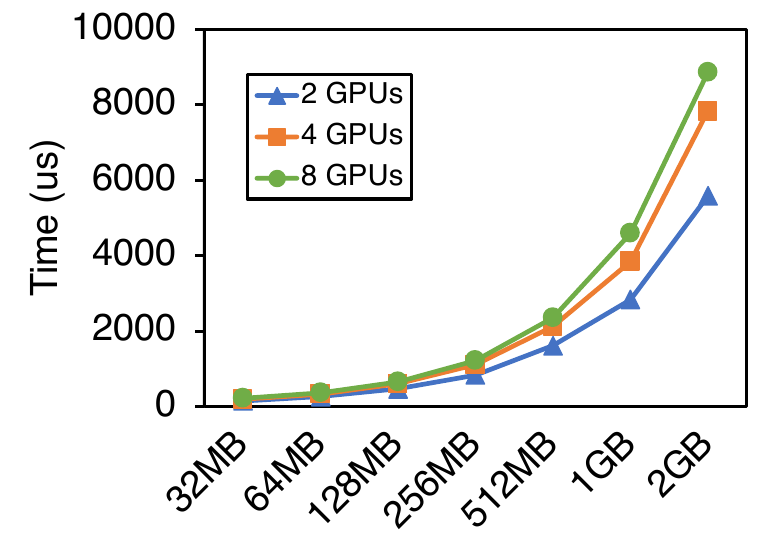}
    \caption{All-to-all\label{fig:a2a_test}}
    \end{subfigure}
\caption{\label{fig:nccl-tests} Time for all-gather and all-to-all collective operations under different message sizes and different number of GPUs using the NCCL Tests on an 8xA100 server.}
\end{figure}

As an example, we profile the all-gather and all-to-all operations on an 8-GPU machine. The total size of the matrix and the corresponding runtime are shown in Figure~\ref{fig:nccl-tests}. We observe that these two factors exhibit a clear log–log relationship, indicating that $\beta$ is determined solely by the type of collective operation and the number of GPUs and is largely independent of the size of the problem. Consequently, we can profile the $\beta$ values once and reuse them in the subsequent analysis. 

\subsection{Automatic Graph Parallelism (AGP)}
\begin{algorithm}[t]
\caption{Automatic Graph Parallelization Algorithm}
\label{alg:auto_gp}
\LinesNumbered
\KwIn{Graph $G$ with $N$ nodes, number of GPUs available $P$}
\KwOut{Graph parallelization strategy $c$, scaling factor $s$}

$B \gets []$; 
measure $t_{\text{iter}}(1)$;
$k \gets \frac{t_{\text{iter}}(1)}{N}$; 
\For{$i \gets 2$ \textbf{to} $P$}{
    \For{$c \in \{\text{GP-AG}, \text{GP-A2A}, ...\}$}{
        measure $b = \beta_c(i)$;
        \If{$\frac{i \cdot b}{i - 1} \leq k$}{ 
            append $\frac{i \cdot b}{i - 1}$ to $B$\;
        }
    }
}
$c, s \gets \argmin_{c,s}(B)$;

\Return{$c, s$};
\end{algorithm}

Based on our formulation, we can estimate the speedup as the system scales and, consequently, determine the “best parallelization strategy”  given a graph dataset and system configuration. Consider scaling the system from $p$ GPUs to $sp$ GPUs. Speedup can only be achieved if the following condition is met: 
\begin{equation}   
t_{iter}(sp) \leq t_{iter}(p). 
\end{equation}
\noindent By substitution, we have:
\begin{equation}   
\alpha(sp) \cdot E + \beta_c(sp) \cdot N \leq \alpha(p) \cdot E + \beta_c(p) \cdot N. 
\end{equation}

\noindent After reorganizing the inequality, we obtain: 

\begin{equation}   
(\beta_c(sp) - \beta_c(p)) \cdot N \leq
\frac{(s-1)\alpha(p)}{s} \cdot E. 
\end{equation}

\noindent $\alpha(p)$ can be further simplified as $\frac{\alpha(1)}{p}$, where $\alpha(1)E$ is the runtime using one GPU, since no communication is needed:

\begin{equation}   
(\beta_c(sp) - \beta_c(p)) \leq
\frac{(s-1)\alpha(1)E}{spN} \simeq \frac{(s-1)t_{iter}(1)}{spN}. 
\end{equation}

Since $t_{iter}(1)$ and $N$ are fixed, the above ratio can be considered as a constant: 
\begin{equation}   
\frac{sp(\beta_c(sp) - \beta_c(p))}{s-1} \leq
\frac{t_{iter}(1)}{N} = k. 
\end{equation}
Therefore, we only need to check if the term on the left side is smaller than the constant $k$. Furthermore, the term $\beta_c(sp) - \beta_c(p)$ determines the speedup: the smaller the term, the greater the speedup achieved by scaling the system. Additionally, consider the special case where $p = 1$; in this case, the inequality becomes:  
\begin{equation}    
\frac{s\beta_c(s)}{s-1} \leq k. 
\end{equation}
By comparing this term across different $c$ and $s$, we can select the optimal graph parallelization strategy. The details are provided in Algorithm~\ref{alg:auto_gp}. {Note that our framework is extensible, i.e., one can incorporate additional parallelization strategies as needed.}

\section{Experimental Evaluation}
We first evaluate the scalability of our distributed training framework. Then, we examine the effectiveness of the automatic parallelization framework across various datasets and system configurations. Finally, we compare the efficiency and accuracy of our method with TorchGT~\cite{10.1109/SC41406.2024.00083}, the state-of-the-art training framework. {We do not report end-to-end speedup compared to DistDGL~\cite{10.1145/3534678.3539177} and TorchGT~\cite{10.1109/SC41406.2024.00083} because DistDGL does not support full-graph training, and TorchGT encounters out-of-memory issues without graph reordering and partitioning. To the best of our knowledge, we are the first to support distributed graph transformer training on full graphs.}

\subsection{Experimental Setup}
We conducted our experiments on two different servers. One is equipped with 8 A100 SXM4 GPUs, where each pair of GPUs is connected by 12 NVLinks\footnote{NVLink is a high-speed, wire-based interconnect technology that enables point-to-point communication between GPUs.}, delivering up to 600 GB/s of bidirectional P2P bandwidth. The other server is equipped with 8 H100 SXM5 GPUs connected by 18 NVLinks, delivering up to 900 GB/s of bidirectional P2P bandwidth.

We implemented our graph parallelization algorithms and distributed graph transformer using PyTorch 2.3.0+cu121~\cite{paszke2017automatic} and DGL 2.3.0+cu121~\cite{wang2019dgl}. For the model design, we set the hidden dimension to 128, following the configuration in~\cite{shirzad2023exphormer}, and set the number of attention heads to 8 to avoid additional communication in GP-A2A. All performance results are {\em averaged} over 10 runs after a 2-run warm-up.

\subsection {Datasets}
We used three real-world benchmark datasets—ogbn-proteins, ogbn-products, and Reddit~\cite{hu2020open, hamilton2017inductive}—to evaluate the scalability and efficiency of sparse graph attention, graph parallelization algorithms, and the automatic optimization framework. We further evaluate the full-graph training accuracy of the graph transformer on two node prediction tasks: ogbn-arxiv and ogbn-products~\cite{hu2020open}. The properties of our graphs are summarized in Table~\ref{tab:microbenchmark}.

\subsection{Performance and Scalability of Automatic Graph Parallelism}
\begin{figure}[t]
  \centering
  \includegraphics[width=0.9\linewidth]{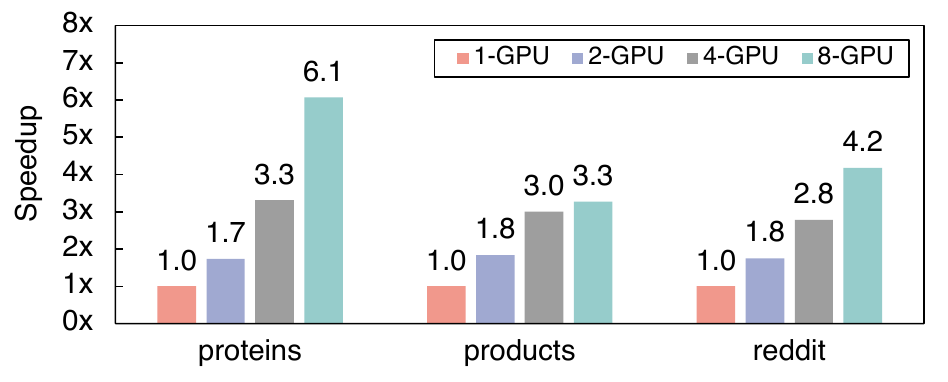}
  \caption{The speedup values when using different number of GPUs on an 8x A100 server across different graphs.}
  \label{fig:a100x8}
\end{figure}

We first measure the iteration {(per-epoch) time speedup} with different number of GPUs to evaluate the scalability. Figure~\ref{fig:a100x8} presents the speedup numbers when training on the A100 server. It can be seen that our “automated selection” of graph parallelism strategy results in consistent and near-linear scalability, achieving up to 6.1x speedup on proteins, 3.3x on products, and 4.2x on reddit when using 8 GPUs, compared to the single-GPU execution. These results clearly demonstrate that our proposed framework effectively exploits parallelism across diverse graph structures.

\begin{figure}[t]
  \centering
  \includegraphics[width=0.9\linewidth]{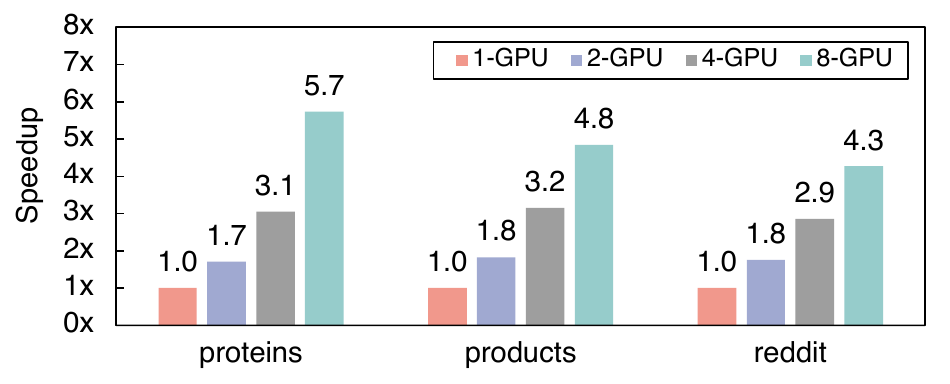}
  \caption{The speedup values when using different number of GPUs on an 8x H100 server across different graphs.}
  \label{fig:h100x8}

  \vspace{-0.1in}
\end{figure}

On the other hand, Figure~\ref{fig:h100x8} presents the speedup values when training on the H100 server. One can observe similar trends as shown in the A100 experiments, suggesting that our method works across different systems. In particular, the H100 server achieves 5.7x, 4.8x, and 4.3x speedup on proteins, products, and reddit, respectively, with 8 GPUs, further confirming the robust scalability and system-agnostic performance of our approach.

\begin{figure}[ht]
\centering
    \begin{subfigure}[b]{0.235\textwidth}
    \centering
    \includegraphics[width=\textwidth]{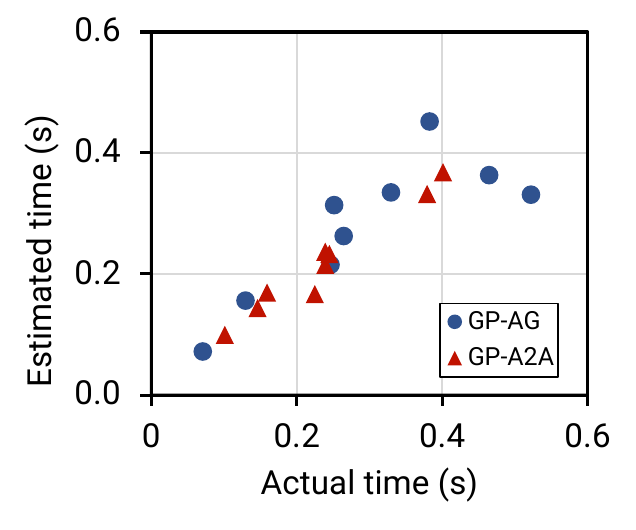}
    \caption{A100}
    \end{subfigure}
    \begin{subfigure}[b]{0.235\textwidth}
    \centering
    \includegraphics[width=\textwidth]{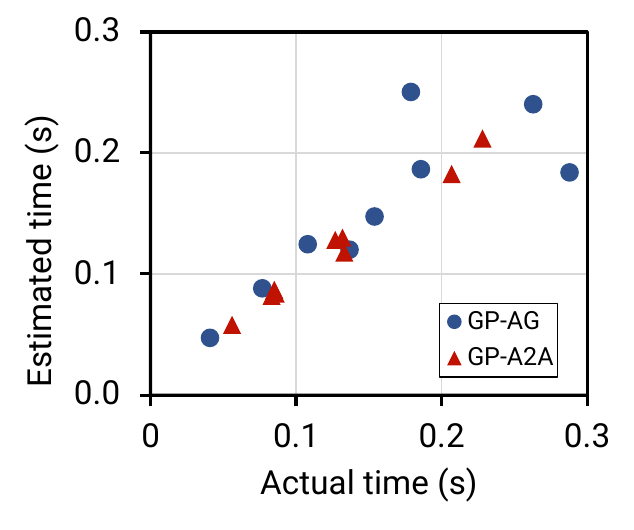}
    \caption{H100}
    \end{subfigure}
\caption{\label{fig:speedup} The relationship between the actual and estimated iteration times under various parallelization strategies and system setups.}
\end{figure}

To evaluate the accuracy of the AGP algorithm’s performance estimation, we also collected the estimated and actual iteration times for different scaling factors, datasets, parallelization strategies, and server configurations. As plotted in Figure~\ref{fig:speedup}, the estimated and actual times exhibit a strong correlation on both A100 and H100 servers, allowing the algorithm to consistently select the optimal strategy.

Overall, one can make two key observations from these experiments:

$\bullet$ The best parallelization strategy varies across different graphs, even when using the same system. For example, GP-AG achieves the best throughput with 8 GPUs on the ogbn-proteins dataset, whereas GP-A2A performs better on the ogbn-products dataset. The optimal strategy depends on both the system and the graph properties, making it essential to adaptively select the best approach to achieve optimal training performance.

$\bullet$ Given a target graph and a target GPU system, with our automatic graph parallelization algorithm, we can determine the best strategy for distributed training across different parallelization schemes, thus avoiding sub-optimal performance that would occur when relying on a single parallelization strategy.  

\subsection{Efficiency of Sparse Graph Attention with Sparse Operations}

\begin{figure}[t]
\centering
    \begin{subfigure}[b]{0.235\textwidth}
    \centering
    \includegraphics[width=\textwidth]{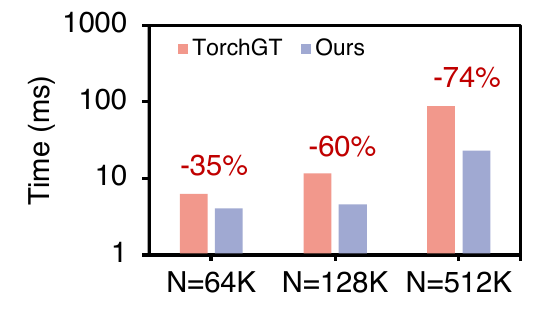}
    \caption{Execution time}
    \end{subfigure}
    \begin{subfigure}[b]{0.235\textwidth}
    \centering
    \includegraphics[width=\textwidth]{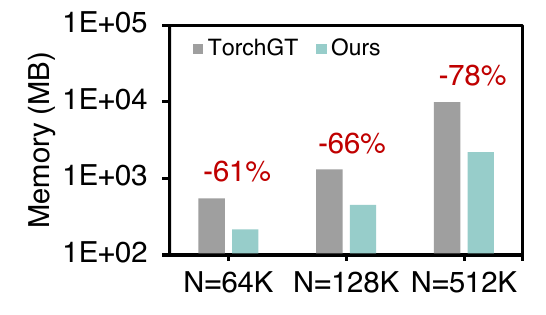}
    \caption{Memory consumption}
    \end{subfigure}
\caption{\label{fig:sparse_n}Comparison of sparse graph attention with TorchGT using different numbers of nodes when $d=128$.}
\end{figure}

\begin{figure}[t]
\centering
    \begin{subfigure}[b]{0.235\textwidth}
    \centering
    \includegraphics[width=\textwidth]{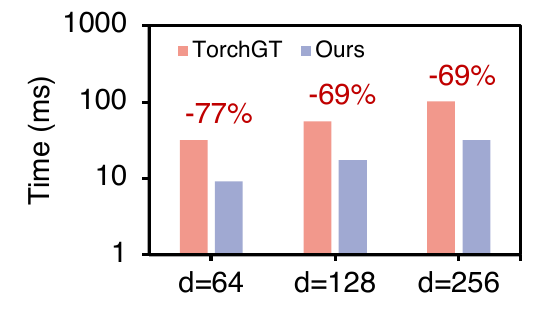}
    \caption{Execution time}
    \end{subfigure}
    \begin{subfigure}[b]{0.235\textwidth}
    \centering
    \includegraphics[width=\textwidth]{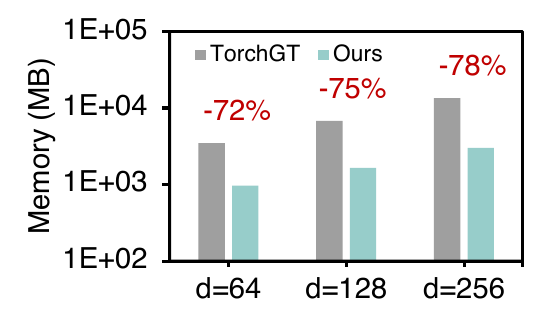}
    \caption{Memory consumption}
    \end{subfigure}
\caption{\label{fig:sparse_h}Comparison of sparse graph attention with TorchGT using different hidden dimension when $N=256K$.}
\end{figure}

We record the time and memory consumption of sparse 
graph attention when training Graph Transformer~\cite{dwivedi2020generalization} on the ogbn-products dataset, following the settings in TorchGT~\cite{10.1109/SC41406.2024.00083}. First, we discuss performance under different graph sizes by setting the number of nodes to 64K, 128K, and 512K. As shown in Figure ~\ref{fig:sparse_n}(a), our implementation of sparse graph attention is much more efficient than TorchGT, and the larger the graph size, the greater the speedup; in fact, we observe a speedup of $3.8\times$ when $N=512K$. In addition to the execution time, we also evaluated the memory efficiency of our implementation. As shown in Figure~\ref{fig:sparse_n}(b), our implementation consistently consumes less memory across different graph sizes, and we reduce memory consumption by $78\%$ when $N=512K$. Second, we discuss performance under different model sizes by setting the hidden dimension to 64, 128, and 256. The results in Figure~\ref{fig:sparse_h} indicate that our proposed method outperforms TorchGT in both execution time and memory consumption, demonstrating consistent improvements across diverse scenarios.

\begin{figure}[t]
  \centering
  \includegraphics[width=0.75\linewidth]{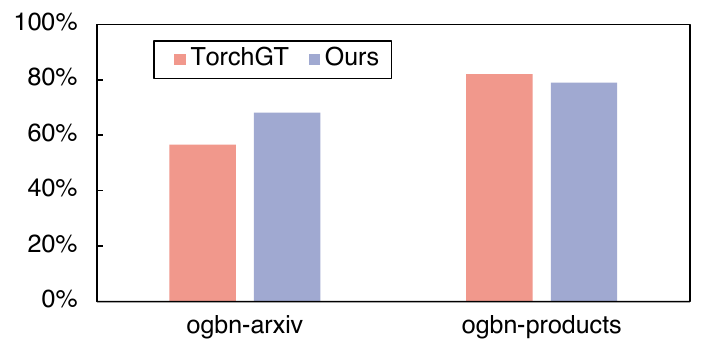}
  \vspace{-0.1in}
  \caption{Comparison of accuracy on different node classification dataset. }
  \label{fig:results}
  \vspace{-0.1in}
\end{figure}

\begin{figure}[t]
  \centering
  \includegraphics[width=0.75\linewidth]{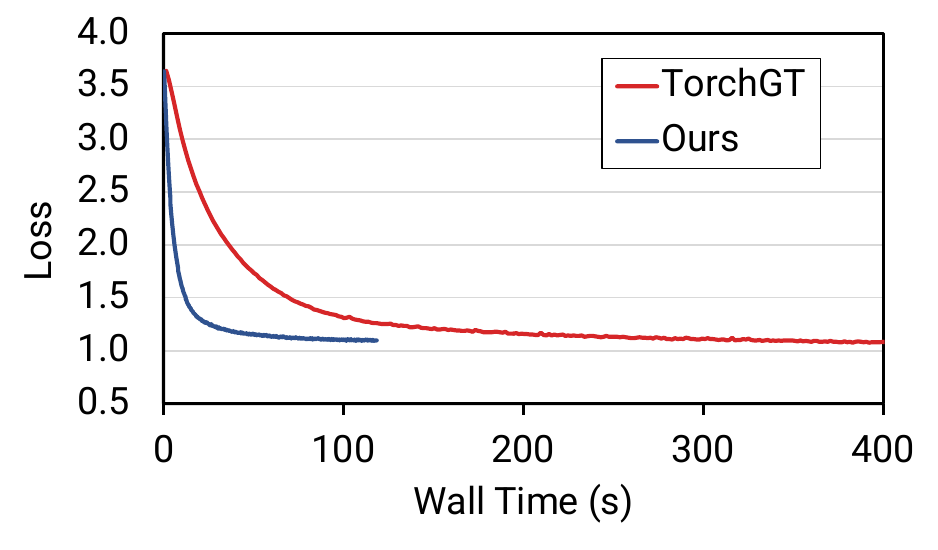}
  \vspace{-0.1in}
  \caption{Comparison of loss versus the wall time on ogbn-products dataset. }
  \label{fig:learning_curve}
  \vspace{-0.1in}
\end{figure}

Regarding the performance of the model, we train a 3-layer Graph Transformer model on different datasets using our framework and TorchGT, with the results compared in Figure~\ref{fig:results}. Our framework achieves model performance similar to that of TorchGT. While the model architectures are identical, TorchGT further applies graph clustering in each epoch to reduce the number of edges and communication between GPUs, which may alter the graph's semantics. We emphasize that our method is orthogonal to graph clustering and can be integrated with TorchGT for further improvements. {Figure~\ref{fig:learning_curve} shows an example. Following TorchGT’s setup for distributed training with 2 GPUs on the ogbn-products dataset, we observe a significant speedup, up to an 83\% reduction in time to reach the same training loss.}

In summary, previous works implement sparse attention by first scattering the \textbf{Q} and \textbf{K} matrices based on edge indices and then computing the dot product. However, such approaches suffer from low computational and memory efficiency because the scattering process requires substantial data movement. In contrast, our method is more efficient in terms of  {\em both} execution time and memory usage, making it feasible to handle even larger graphs while providing better throughput and scalability.

\subsection{Discussion on Scaling to Larger Systems}
Although our evaluation is limited to single-node systems with up to 8 GPUs, the underlying design of the parallelization strategies, particularly the separation of sparse computation (edge-dependent) and communication (node-dependent) components, suggests that the proposed approach should extend to larger GPU counts. Both GP-AG and GP-A2A rely on standard collective primitives whose scaling behavior has been well-characterized by prior research, and the AGP model explicitly incorporates communication coefficients ($\beta$) that remain valid as the system scales. As a result, the framework is expected to maintain similar efficiency trends on multi-node or higher-GPU configurations, provided that the interconnect bandwidth scales proportionally.

\section{Conclusion}
Graph transformers emerged as a powerful paradigm for learning on graph-structured data, yet their scalability has been hindered by the computational and memory demands of sparse attention. Motivated by this, in this paper, we addressed the challenges behind “distributed graph transformer training” by developing two parallelization strategies, \textbf{GP-AG} and \textbf{GP-A2A}, and designing the adaptive algorithm \textbf{AGP} to automatically select the best strategy based on the characteristics of the target graph and the target hardware system. Our framework leverages sparse operators, reduces memory consumption, and achieves substantial improvements in training throughput on {large graph} benchmarks. These results demonstrate that distributed graph transformer training can be made {\em both} efficient and practical, advancing their potential as graph foundation models.  

\bibliography{refs}

@article{Zhuo2025,
  title={A Closer Look at Graph Transformers: Cross-Aggregation and Beyond},
  author={Jiaming Zhuo and Ziyi Ma and Yintong Lu and Yuwei Liu and Kun Fu and Di Jin and Chuan Wang and Wu Wenning and Zhen Wang and Xiaochun Cao and Liang Yang},
  journal={Advances in Neural Information Processing Systems (to appear)},
  volume={},
  year={2025}}

@article{rampavsek2022recipe,
  title={Recipe for a general, powerful, scalable graph transformer},
  author={Ramp{\'a}{\v{s}}ek, Ladislav and Galkin, Michael and Dwivedi, Vijay Prakash and Luu, Anh Tuan and Wolf, Guy and Beaini, Dominique},
  journal={Advances in Neural Information Processing Systems},
  volume={35},
  pages={14501--14515},
  year={2022}
}

@inproceedings{shirzad2023exphormer,
  title={Exphormer: Sparse transformers for graphs},
  author={Shirzad, Hamed and Velingker, Ameya and Venkatachalam, Balaji and Sutherland, Danica J and Sinop, Ali Kemal},
  booktitle={International Conference on Machine Learning},
  eprinttype={arXiv}, eprint={2303.06147},
  year={2023}
}

@inproceedings{
wu2023simplifying,
title={Simplifying and Empowering Transformers for Large-Graph Representations},
author={Qitian Wu and Wentao Zhao and Chenxiao Yang and Hengrui Zhang and Fan Nie and Haitian Jiang and Yatao Bian and Junchi Yan},
booktitle={Thirty-seventh Conference on Neural Information Processing Systems},
year={2023},
url={https://openreview.net/forum?id=R4xpvDTWkV}
}

@article{dubey2024llama,
  title={The llama 3 herd of models},
  author={Dubey, Abhimanyu and Jauhri, Abhinav and Pandey, Abhinav and Kadian, Abhishek and Al-Dahle, Ahmad and Letman, Aiesha and Mathur, Akhil and Schelten, Alan and Yang, Amy and Fan, Angela and others},
  journal={arXiv preprint arXiv:2407.21783},
  year={2024}
}

@article{dwivedi2020generalization,
  title={A generalization of transformer networks to graphs},
  author={Dwivedi, Vijay Prakash and Bresson, Xavier},
  journal={arXiv preprint arXiv:2012.09699},
  year={2020}
}

@article{hu2020open,
  title={Open graph benchmark: Datasets for machine learning on graphs},
  author={Hu, Weihua and Fey, Matthias and Zitnik, Marinka and Dong, Yuxiao and Ren, Hongyu and Liu, Bowen and Catasta, Michele and Leskovec, Jure},
  journal={Advances in neural information processing systems},
  volume={33},
  pages={22118--22133},
  year={2020}
}

@article{wang2019dgl,
    title={Deep Graph Library: A Graph-Centric, Highly-Performant Package for Graph Neural Networks},
    author={Minjie Wang and Da Zheng and Zihao Ye and Quan Gan and Mufei Li and Xiang Song and Jinjing Zhou and Chao Ma and Lingfan Yu and Yu Gai and Tianjun Xiao and Tong He and George Karypis and Jinyang Li and Zheng Zhang},
    year={2019},
    journal={arXiv preprint arXiv:1909.01315}
}

@article{paszke2017automatic,
  title={Automatic differentiation in PyTorch},
  author={Paszke, Adam and Gross, Sam and Chintala, Soumith and Chanan, Gregory and Yang, Edward and DeVito, Zachary and Lin, Zeming and Desmaison, Alban and Antiga, Luca and Lerer, Adam},
  year={2017}
}

@misc{nccl-tests,
  author    = {NVIDIA},
  title     = {NCCL Tests},
  year      = {2024},
  url       = {https://github.com/NVIDIA/nccl-tests},
  note      = {Accessed: 2024-10-10}
}

@inproceedings{hoffmann2022training,
  title={Training compute-optimal large language models},
  author={Hoffmann, Jordan and Borgeaud, Sebastian and Mensch, Arthur and Buchatskaya, Elena and Cai, Trevor and Rutherford, Eliza and de Las Casas, Diego and Hendricks, Lisa Anne and Welbl, Johannes and Clark, Aidan and others},
  booktitle={Proceedings of the 36th International Conference on Neural Information Processing Systems},
  pages={30016--30030},
  year={2022}
}

@article{achiam2023gpt,
  title={Gpt-4 technical report},
  author={Achiam, Josh and Adler, Steven and Agarwal, Sandhini and Ahmad, Lama and Akkaya, Ilge and Aleman, Florencia Leoni and Almeida, Diogo and Altenschmidt, Janko and Altman, Sam and Anadkat, Shyamal and others},
  journal={arXiv preprint arXiv:2303.08774},
  year={2023}
}

@article{comanici2025gemini,
  title={Gemini 2.5: Pushing the frontier with advanced reasoning, multimodality, long context, and next generation agentic capabilities},
  author={Comanici, Gheorghe and Bieber, Eric and Schaekermann, Mike and Pasupat, Ice and Sachdeva, Noveen and Dhillon, Inderjit and Blistein, Marcel and Ram, Ori and Zhang, Dan and Rosen, Evan and others},
  journal={arXiv preprint arXiv:2507.06261},
  year={2025}
}

@inproceedings{gui2024g,
  title={G-adapter: Towards structure-aware parameter-efficient transfer learning for graph transformer networks},
  author={Gui, Anchun and Ye, Jinqiang and Xiao, Han},
  booktitle={Proceedings of the AAAI Conference on Artificial Intelligence},
  volume={38},
  number={11},
  pages={12226--12234},
  year={2024}
}

@inproceedings{sun2022gppt,
  title={Gppt: Graph pre-training and prompt tuning to generalize graph neural networks},
  author={Sun, Mingchen and Zhou, Kaixiong and He, Xin and Wang, Ying and Wang, Xin},
  booktitle={Proceedings of the 28th ACM SIGKDD Conference on Knowledge Discovery and Data Mining},
  pages={1717--1727},
  year={2022}
}

@inproceedings{
liu2024one,
title={One For All: Towards Training One Graph Model For All Classification Tasks},
author={Hao Liu and Jiarui Feng and Lecheng Kong and Ningyue Liang and Dacheng Tao and Yixin Chen and Muhan Zhang},
booktitle={The Twelfth International Conference on Learning Representations},
year={2024},
url={https://openreview.net/forum?id=4IT2pgc9v6}
}

@inproceedings{shi2021masked,
  title={Masked Label Prediction: Unified Message Passing Model for Semi-Supervised Classification},
  author={Shi, Yunsheng and Huang, Zhengjie and Feng, Shikun and Zhong, Hui and Wang, Wenjing and Sun, Yu},
  booktitle={Proceedings of the Thirtieth International Joint Conference on Artificial Intelligence},
  pages={1548--1554},
  year={2021},
  organization={International Joint Conferences on Artificial Intelligence Organization}
}

@article{zi2024prog,
  title={Prog: A graph prompt learning benchmark},
  author={Zi, Chenyi and Zhao, Haihong and Sun, Xiangguo and Lin, Yiqing and Cheng, Hong and Li, Jia},
  journal={Advances in Neural Information Processing Systems},
  volume={37},
  pages={95406--95437},
  year={2024}
}

@article{liu2025graph,
  title={Graph foundation models: Concepts, opportunities and challenges},
  author={Liu, Jiawei and Yang, Cheng and Lu, Zhiyuan and Chen, Junze and Li, Yibo and Zhang, Mengmei and Bai, Ting and Fang, Yuan and Sun, Lichao and Yu, Philip S and others},
  journal={IEEE Transactions on Pattern Analysis and Machine Intelligence},
  year={2025},
  publisher={IEEE}
}

@inproceedings{10.1109/SC41406.2024.00083,
author = {Zhang, Meng and Sun, Jie and Hu, Qinghao and Sun, Peng and Wang, Zeke and Wen, Yonggang and Zhang, Tianwei},
title = {TorchGT: A Holistic System for Large-Scale Graph Transformer Training},
year = {2024},
isbn = {9798350352917},
publisher = {IEEE Press},
url = {https://doi.org/10.1109/SC41406.2024.00083},
doi = {10.1109/SC41406.2024.00083},
booktitle = {Proceedings of the International Conference for High Performance Computing, Networking, Storage, and Analysis},
articleno = {77},
numpages = {17},
location = {Atlanta, GA, USA},
series = {SC '24}
}

@INPROCEEDINGS{8638042,
  author={Nisa, Israt and Sukumaran-Rajam, Aravind and Kurt, Sureyya Emre and Hong, Changwan and Sadayappan, P.},
  booktitle={2018 IEEE 25th International Conference on High Performance Computing (HiPC)}, 
  title={Sampled Dense Matrix Multiplication for High-Performance Machine Learning}, 
  year={2018},
  volume={},
  number={},
  pages={32-41},
  doi={10.1109/HiPC.2018.00013}}

@inproceedings{rahman2021fusedmm,
  title={Fusedmm: A unified sddmm-spmm kernel for graph embedding and graph neural networks},
  author={Rahman, Md Khaledur and Sujon, Majedul Haque and Azad, Ariful},
  booktitle={2021 IEEE International Parallel and Distributed Processing Symposium (IPDPS)},
  pages={256--266},
  year={2021},
  organization={IEEE}
}

@article{chan2007collective,
  title={Collective communication: theory, practice, and experience},
  author={Chan, Ernie and Heimlich, Marcel and Purkayastha, Avi and Van De Geijn, Robert},
  journal={Concurrency and Computation: Practice and Experience},
  volume={19},
  number={13},
  pages={1749--1783},
  year={2007},
  publisher={Wiley Online Library}
}

@techreport{saad1990sparskit,
  title={SPARSKIT: A basic tool kit for sparse matrix computations},
  author={Saad, Youcef},
  year={1990}
}

@inproceedings{zheng2020distdgl,
  title={DistDGL: Distributed graph neural network training for billion-scale graphs},
  author={Zheng, Da and Ma, Chao and Wang, Minjie and Zhou, Jinjing and Su, Qidong and Song, Xiang and Gan, Quan and Zhang, Zheng and Karypis, George},
  booktitle={2020 IEEE/ACM 10th Workshop on Irregular Applications: Architectures and Algorithms (IA3)},
  pages={36--44},
  year={2020},
  organization={IEEE}
}

@inproceedings{narayanan2021efficient,
  title={Efficient large-scale language model training on gpu clusters using megatron-lm},
  author={Narayanan, Deepak and Shoeybi, Mohammad and Casper, Jared and LeGresley, Patrick and Patwary, Mostofa and Korthikanti, Vijay and Vainbrand, Dmitri and Kashinkunti, Prethvi and Bernauer, Julie and Catanzaro, Bryan and others},
  booktitle={Proceedings of the international conference for high performance computing, networking, storage and analysis},
  pages={1--15},
  year={2021}
}

@article{hamilton2017inductive,
  title={Inductive representation learning on large graphs},
  author={Hamilton, Will and Ying, Zhitao and Leskovec, Jure},
  journal={Advances in neural information processing systems},
  volume={30},
  year={2017}
}

@article{10.14778/3717755.3717776,
author = {Bajaj, Saurabh and Son, Hojae and Liu, Juelin and Guan, Hui and Serafini, Marco},
title = {Graph Neural Network Training Systems: A Performance Comparison of Full-Graph and Mini-Batch},
year = {2024},
issue_date = {December 2024},
publisher = {VLDB Endowment},
volume = {18},
number = {4},
issn = {2150-8097},
url = {https://doi.org/10.14778/3717755.3717776},
doi = {10.14778/3717755.3717776},
journal = {Proc. VLDB Endow.},
month = dec,
pages = {1196–1209},
numpages = {14}
}

@inproceedings{10.1145/3534678.3539177,
author = {Zheng, Da and Song, Xiang and Yang, Chengru and LaSalle, Dominique and Karypis, George},
title = {Distributed Hybrid CPU and GPU training for Graph Neural Networks on Billion-Scale Heterogeneous Graphs},
year = {2022},
isbn = {9781450393850},
publisher = {Association for Computing Machinery},
address = {New York, NY, USA},
url = {https://doi.org/10.1145/3534678.3539177},
doi = {10.1145/3534678.3539177},
booktitle = {Proceedings of the 28th ACM SIGKDD Conference on Knowledge Discovery and Data Mining},
pages = {4582–4591},
numpages = {10},
location = {Washington DC, USA},
series = {KDD '22}
}

@inproceedings {285052,
author = {Tianfeng Liu and Yangrui Chen and Dan Li and Chuan Wu and Yibo Zhu and Jun He and Yanghua Peng and Hongzheng Chen and Hongzhi Chen and Chuanxiong Guo},
title = {{BGL}: {GPU-Efficient} {GNN} Training by Optimizing Graph Data {I/O} and Preprocessing},
booktitle = {20th USENIX Symposium on Networked Systems Design and Implementation (NSDI 23)},
year = {2023},
isbn = {978-1-939133-33-5},
address = {Boston, MA},
pages = {103--118},
url = {https://www.usenix.org/conference/nsdi23/presentation/liu-tianfeng},
publisher = {USENIX Association},
month = apr
}

@article{wan2022bns,
  title={{BNS-GCN}: Efficient Full-graph Training of Graph Convolutional Networks with Partition-parallelism and Random Boundary Node Sampling},
  author={Wan, Cheng and Li, Youjie and Li, Ang and Kim, Nam Sung and Lin, Yingyan},
  journal={Proceedings of Machine Learning and Systems},
  volume={4},
  pages={673--693},
  year={2022}
}

@article{wan2023scalable,
  title={Scalable and efficient full-graph gnn training for large graphs},
  author={Wan, Xinchen and Xu, Kaiqiang and Liao, Xudong and Jin, Yilun and Chen, Kai and Jin, Xin},
  journal={Proceedings of the ACM on Management of Data},
  volume={1},
  number={2},
  pages={1--23},
  year={2023},
  publisher={ACM New York, NY, USA}
}

@article{10.14778/3725688.3725700,
author = {Fu, Zhenbo and Ai, Xin and Wang, Qiange and Zhang, Yanfeng and Lu, Shizhan and Chen, Chaoyi and Cao, Chunyu and Yuan, Hao and Wei, Zhewei and Gu, Yu and Wen, Yingyou and Yu, Ge},
title = {NeutronTask: Scalable and Efficient Multi-GPU GNN Training with Task Parallelism},
year = {2025},
issue_date = {February 2025},
publisher = {VLDB Endowment},
volume = {18},
number = {6},
issn = {2150-8097},
url = {https://doi.org/10.14778/3725688.3725700},
doi = {10.14778/3725688.3725700},
journal = {Proc. VLDB Endow.},
month = feb,
pages = {1705–1719},
numpages = {15}
}

\end{document}